\documentclass[psfig]{aa}
\usepackage{graphics}
\usepackage{psfig}

\begin{document}

\title {Multiwavelength optical observations of chromospherically
active binary systems}
\subtitle{IV. The X-ray/EUV selected binary
BK Psc (2RE J0039+103)
\thanks{Based on observations made
with the 2.2~m telescope of the German-Spanish Astronomical Centre,
Calar Alto (Almer\'{\i}a, Spain),
operated by the Max-Planck-Institute for Astronomy,
Heidelberg, jointly with the Spanish National Commission for Astronomy,
with the Nordic Optical Telescope (NOT),
 operated on the island of La Palma jointly by Denmark, Finland,
 Iceland, Norway and Sweden, in the Spanish Observatorio del
 Roque de Los Muchachos of the Instituto de Astrof\'{\i}sica de Canarias,
and with the Isaac Newton Telescope (INT)
 operated on the island of La Palma by the Isaac Newton Group in
 the Spanish Observatorio del Roque de Los Muchachos of the
 Instituto de Astrof\'{\i}sica de Canarias. }
}

\titlerunning{The X-ray/EUV selected binary
BK Psc} 

\author{
M.C.~G\'alvez
\and D.~Montes
\and M.J.~Fern\'{a}ndez-Figueroa
\and J. L\'opez-Santiago
\and E.~De Castro
\and M.~Cornide
}

\offprints{ D.~Montes}
\mail{dmg@astrax.fis.ucm.es}

\institute{
Departamento de Astrof\'{\i}sica,
Facultad de Ciencias F\'{\i}sicas,
 Universidad Complutense de Madrid, E-28040 Madrid, Spain\\
e-mail: {\tt dmg@astrax.fis.ucm.es}
}

\date{Received 11 February 2002 / Accepted 16 April 2002}

\abstract{
We present high resolution echelle spectra taken during
four observing runs from 1999 to 2001 of the recently
X-ray/EUV selected chromospherically active binary 
BK Psc (2RE J0039+103).
Our observations confirm the single-lined spectroscopic binary (SB1) nature
of this system and allow us to obtain, for the first time,
the orbital solution of the system as in the case of a SB2 system.
We have determined precise radial velocities of both components: 
for the primary by using the cross correlation technique, 
and for the secondary by using its chromospheric emission lines.
We have obtained a circular orbit with an orbital period
of 2.1663 days, very close to its photometric period of 2.24 days
(indicating synchronous rotation).
The spectral type (K5V) we determined for our spectra and the mass ratio
(1.8) and minimum masses ($Msin^{3}i$) resulting from the orbital
solution are compatible with the observed K5V primary
and an unseen M3V secondary.
Using this spectral classification, the projected rotational
velocity ($v\sin{i}$, of 17.1 km s$^{-1}$)
obtained from the width of the cross-correlation function 
and the data provided by {\sc Hipparcos}, we have derived other
fundamental stellar parameters. 
The kinematics and the non- detection of the Li~{\sc i} line indicate
that it is an old star.
The analysis of the optical chromospheric activity indicators
from the Ca~{\sc ii} H \& K to Ca~{\sc ii} IRT lines,
by using the spectral subtraction technique, indicates that
both components of the binary system show
high levels of chromospheric activity.
H$\alpha$ emission above the continuum from both components is a 
persistent feature of this system during
the period 1999 to 2001 of our observations as well as in previous 
observations.  
The H$\alpha$ and H$\beta$ emission seems
to arise from prominence-like material, and the Ca~{\sc ii} IRT emission
from plage-like regions.
}

\maketitle

\begin{table*}
\caption[]{Observing log
\label{tab:obslog}}
\begin{flushleft}
\small
\begin{tabular}{ccccccccccccccccccc}
\noalign{\smallskip}
\hline
\noalign{\smallskip}
\multicolumn{4}{c}{2.2~m-FOCES 1999/07} &\ &
\multicolumn{4}{c}{INT-MUSICOS 2000/08} &\ &
\multicolumn{4}{c}{NOT-SOFIN 2000/11} &\ &
\multicolumn{4}{c}{2.2~m-FOCES 2001/09} \\
\cline{1-4}\cline{6-9}\cline{11-14}\cline{16-19}
\noalign{\smallskip}
\scriptsize Day & \small UT& \small Exp & \small S/N&\ &
\scriptsize Day & \small UT& \small Exp & \small S/N&\ &
\scriptsize Day & \small UT& \small Exp & \small S/N&\ &
\scriptsize Day & \small UT& \small Exp & \small S/N  
\\
& & \small (s) & \small H$\alpha$ & \ &
& & \small (s) & \small H$\alpha$ & \ &
& & \small (s) & \small H$\alpha$ & \ &
& & \small (s) & \small H$\alpha$  
\scriptsize
\\
\noalign{\smallskip}
\hline
\noalign{\smallskip}
\noalign{\smallskip}
 26 & 02:45 & 2400 & 87 & & 11 & 04:19 & 4000 & 36 & & 6 & 01:03 & 3600 & 84 & & 24 & 01:17 &1600 & 32 \\
 28 & 03:36 & 2000 & 68 & & 12 & 04:14 & 4000 & 64 & & 6 & 23:06 & 3600 & 79 & & 24 & 22:45 &1600 & 57 \\
 30 & 03:11 & 2000 & 75 & & 13 & 04:39 & 4000 & 62 & & 8 & 00:28 & 3600 & 88 & &    &       &     &    \\
    &       &      &    & & 14 & 04:36 & 3600 & 49 & & 8 & 23:30 & 3600 & 96 & &    &       &     &    \\
\noalign{\smallskip}
\noalign{\smallskip}
\hline
\end{tabular}
\end{flushleft}
\end{table*}

\section{Introduction}

This paper is a part of our ongoing project on
multiwavelength optical observations
aimed at studying the chromosphere of active binary systems.
For this purpose we use the information provided by several optical 
spectroscopic features that are formed at different heights in the 
chromosphere
(see Montes et al. 1997, Paper~I; Montes et al. 1998, Paper~II; 
Montes et al. 2000, Paper~III). 
In addition to study stellar activity, the high resolution 
spectroscopic observations we use in this project allow us to
determine radial velocities and to obtain and improve 
fundamental stellar parameters.

In this paper we focus our attention on the X-ray/EUV selected 
chromospherically active binary 
BK Psc (2RE J003939+103925, BD+09 73, G 1-10, LHS 1118).
It is a high proper-motion star with photometry reported by
Stephenson (1986), Sandage \& Kowal (1986) and Weis (1991)
($V$=10.5; $U-B$ = 0.97; $B-V$ = 1.17; $V-R$ = 0.73; $R-I$ = 0.60).
Bidelman (1985) gives a K5 spectral type for this star that was 
confirmed later by Jeffries et al. (1995), but
 Stephenson (1986) listed it as K4:p.
Due to a blue excess in the ($U-B$) color index the presence of a 
white dwarf companion
has been suggested by Weis (1991) and Cutispoto et al. (1999).
Radial velocity variations 
indicate that it is a binary system 
(Jeffries et al. 1995; Cutispoto et al. 1999), 
but no orbital solution has been determined until now.
BK Psc was detected as an extreme ultraviolet (EUV) source
by the {\sc ROSAT} Wide Field Camera all-sky survey
(Pounds et al. 1993; Pye et al. 1995).
The chromospheric activity of this star was detected in
the optical identification program of {\sc ROSAT} EUV Sources 
by Mason et al. (1995) 
($EW$(H$\alpha$)=1.1 \AA\ and $EW$(Ca{\sc ii} K)=2.7 \AA)
and Jeffries et al. (1995)
(strong H$\alpha$ emission above the continuum, $EW$(H$\alpha$)=1.0 \AA).
Finally, Cutispoto et al. (1999) found that the best fit for 
their observed colors
($V$=10.43; $U-B$ = 0.92; $B-V$ = 1.16; $V-R$ = 0.73; $V-I$ = 1.49) and
{\sc Hipparcos} distance ($d=32.8$ pc) is obtained by assuming a 
K5V or K6V primary, a M4V secondary and a possible white dwarf (WD).
%
These authors also confirm the optical variability of this star
(photometric period $P_{\rm phot}=2.24\pm0.04$ days),
previously reported in the SAAO Annual report (1993)
($P_{\rm phot}=2.17$ days).

In this paper we present high resolution echelle spectra of BK Psc,
obtained at different epochs, that allow us to measure 
radial and rotational velocities by using the cross-correlation technique.
With these observations we confirm the binary nature (SB1) of this system
and determine its orbital solution for the first time.
%
%
We have obtained an orbital period of 2.17 days, very close
to the photometric period of 2.24 days, indicating nearly 
synchronous rotation.
Furthermore, we have applied the spectral
subtraction technique to study the chromospheric excess emission 
in the Ca~{\sc ii} H \& K, Ca~{\sc ii} IRT, H$\alpha$ and
other Balmer lines from the primary and secondary components of the system.
Preliminary results for this system can be found in 
G\'alvez et al. (2001) and Montes et al. (2001a). 

In Sect.~2 we give the details of our observations and data reduction.
In Sect.~3 the procedures to obtain the stellar parameters and the 
orbital determination of the binary system are described in more detail 
and the results are discussed.
The individual behavior of the different chromospheric
activity indicators is described in Sect.~4.
Finally in Sect.~5 we give the conclusions.

\begin{table*}
\caption[]{Stellar parameters of \object{BK Psc}
\label{tab:par}}
\begin{flushleft}
\small
\begin{tabular}{l c c c  c l l c c c c }
\noalign{\smallskip}
\hline
\noalign{\smallskip}
  {T$_{\rm sp}$} & {SB} & V &
 {$B-V$} & {$V-R$} & {$P_{\rm orb}^{1}$} & {$P_{\rm phot}$} &
$v\sin{i}^{1}$ & 
$\pi$ & $\mu$$_{\alpha}$ cos $\delta$ & $\mu$$_{\delta}$ \\
             &     &    &      &   & \small (days) & \small (days) &
\small (km s$^{-1}$) &
(mas) & (mas/yr) & (mas/yr) \\
\noalign{\smallskip}
\hline
\noalign{\smallskip}
 K5/6:V/M4:V~+~{\scriptsize WD} & 1 & 10.43 & 1.16 & 0.73   & 2.1663 & 2.24  & 
17.1 & 30.52$\pm$1.79 & 524.9$\pm$2.1 & -198.0$\pm$2.1  \\
\noalign{\smallskip}
\hline
\noalign{\smallskip}
\end{tabular}

$^{1}$ values determined in this paper

\end{flushleft}
\end{table*}

\section{Observations and data reduction}

The spectroscopic observations of BK Psc  were obtained during 
four observing runs:

{1)} {2.2~m-FOCES 1999/07} \\
This run took place on 24-29 July 1999 using the 2.2~m 
telescope at the German Spanish Astronomical
Observatory (CAHA) (Almer\'{\i}a, Spain).
The Fibre Optics Cassegrain Echelle Spectrograph (FOCES)
(Pfeiffer et al. 1998)
was used with a 2048x2048 (15$\mu$) LORAL$\#$11i CCD detector.
The wavelength range covers from
3910 to 9075 \AA\ in 84 orders.
The reciprocal dispersion ranges from 0.03 to 0.07 \AA/pixel
and the spectral resolution,
determined as the full width at half maximum (FWHM)
of the arc comparison lines, ranges from 0.09 to 0.26 \AA.

{2)} {INT-MUSICOS 2000/08} \\
This run was done on 11-14 August 2000, with the 2.5~m 
Isaac Newton telescope (INT) at the
Observatorio del Roque de Los Muchachos (La Palma, Spain) using the
ESA MUSICOS Spectrograph. This is a fibre-fed cross-dispersed echelle
spectrograph (Baudrand \& Bohm 1992) and developed as a part of 
MUlti-SIte COntinuous Spectroscopy (MUSICOS) project. 
During this observing run, a 2148x4148 pixel EEV10a (EEV4280) 
CCD detector was used.
The wavelength range covers from 4000 to 10000 \AA\ in 87 orders. The 
reciprocal dispersion ranges from 0.03 to 0.08 \AA/pixel 
and the spectral resolution (FWHM) from 0.06 to 0.24 \AA.

{3)} {NOT-SOFIN 2000/11}  \\
This run took place on 6-9 November 2000 using the 2.56~m 
Nordic Optical Telescope (NOT).
The Soviet Finnish High Resolution Echelle Spectrograph (SOFIN) was
used with an echelle grating (79 grooves/mm), camera ASTROMED-3200 and a 
1152x298 pixel EEV P88100 CCD detector. The wavelength range covers from 
3900 to 8400 \AA\ in 37 orders. 
The reciprocal dispersion ranges from 0.06 to 0.14 \AA/pixel 
and the spectral resolution (FWHM) from 0.15 to 0.42 \AA.

{4)} {2.2~m-FOCES 2001/09} \\
The last run took place on 21-24 September 2001 using the 2.2~m telescope
 at the German Spanish Astronomical
Observatory (CAHA) (Almer\'{\i}a, Spain).
In this run, the FOCES spectrograph (Pfeiffer et al. 1998)
was used with a 2048x2048 (24$\mu$) SITE$\#$1d CCD detector.
The wavelength range covers from
3450 to 10700 \AA\ in 112 orders.
The reciprocal dispersion ranges from 0.04 to 0.13 \AA/pixel
and the spectral resolution (FWHM) ranges from 0.1 to 0.34 \AA.

In Table~\ref{tab:obslog} we give the observing log.
For each observation we list date (Day), universal time (UT), 
exposure time (Exp), and the signal to noise ratio (S/N) measured 
in the H$\alpha$ line region.

The spectra have been extracted using the standard
reduction procedures in the
IRAF\footnote{IRAF is distributed by the National Optical Observatory,
which is operated by the Association of Universities for Research in
Astronomy, Inc., under contract with the National Science Foundation.}
 package.
The wavelength calibration was obtained by taking
spectra of a Th-Ar lamp.
Finally, the spectra have been normalized by
a polynomial fit to the observed continuum.

\section{Stellar parameters of the binary system}

The adopted stellar  parameters of BK Psc are given in Table~\ref{tab:par}.
Spectral type and the photometric data 
($V$, $B-V$, $V-R$, $P_{\rm phot}$) are taken from Cutispoto et al. (1999).
Orbital period ($P_{\rm orb}$) and projected rotational velocity
($v\sin{i}$) have been determined in this paper (see below).
The astrometric data (parallax, $\pi$, proper motions,  
$\mu$$_{\alpha}$$\cos{\delta}$ and  $\mu$$_{\delta}$) 
are from Hipparcos (ESA 1997) 
and Tycho-2 (H$\o$g et al. 2000) catalogues.

\subsection{Spectral types}


In order to obtain the spectral type of this binary sytem we have 
compared our high resolution echelle spectra, in several spectral 
orders free of lines sensitive to chromospheric activity, with 
spectra of inactive reference stars of different spectral types and 
luminosity classes observed during the same observing run.
This analysis made use of the program {\sc starmod} 
developed at Penn State University (Barden 1985) and modified later by us.
With this program a synthesized stellar spectrum is constructed
from artificially rotationally broadened, radial-velocity
shifted, and weighted spectra of appropiate reference stars.
We have obtained the best fit between observed and synthetic spectra
when we use a K5V primary component without any contribution from 
a secondary component. 
As reference stars we have used the K5V stars 
HD~ 154363 for the first run,
61 Cyg A for for the other three runs.
This spectral classification is in agreement with 
K5 spectral type reported by Bidelman (1985) and Jeffries et al. (1995),
and the K5/6:V~+~M4:V classification given by Cutispoto et al. (1999), 
where the M4V secondary has no apreciable contribution to the spectra.

\begin{figure*}
{\psfig{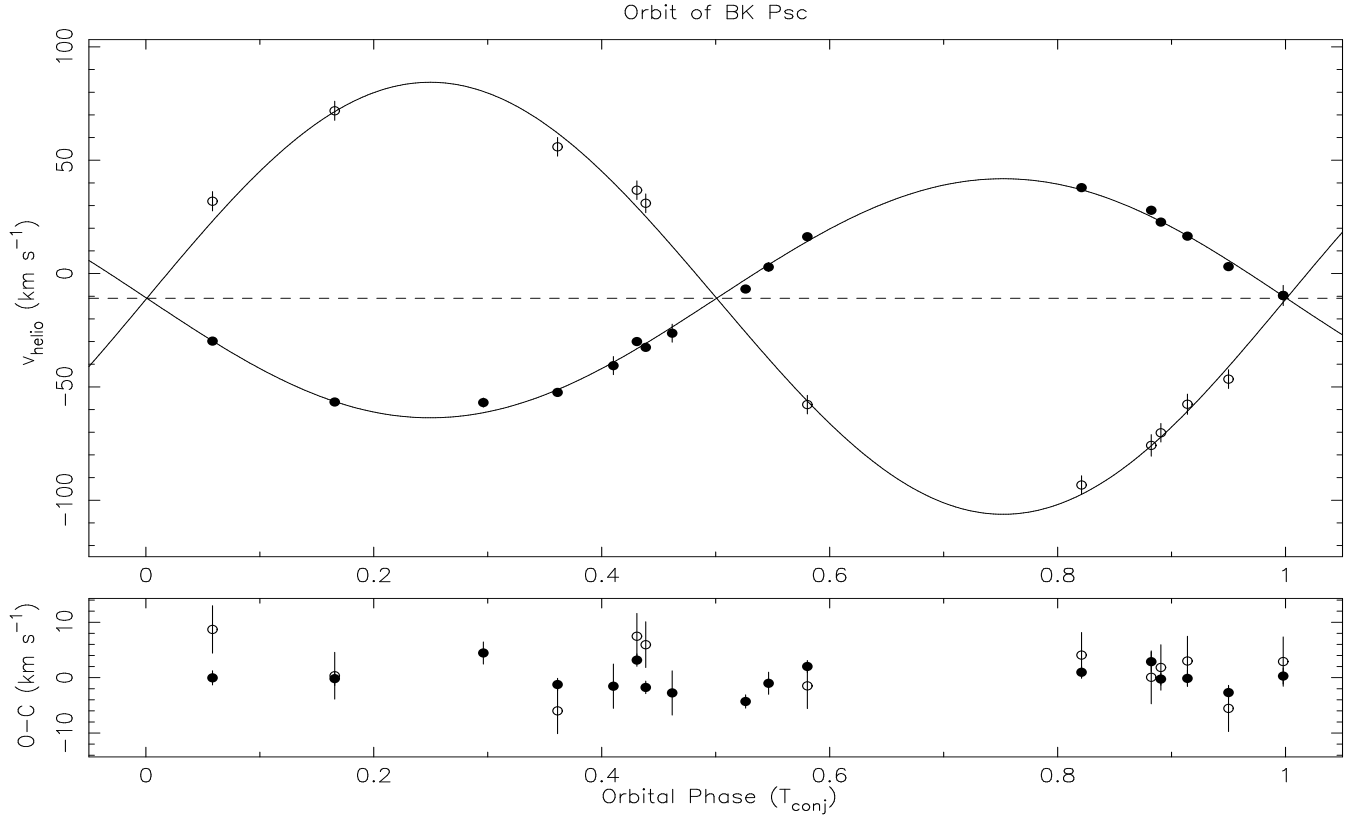}}
\caption[ ]{Radial velocity data and fit vs the orbital phase.
Solid circles represent the primary and open circles represent the secondary.
The solid curves represent a minimum $\chi^{2}$ fit orbit solution
as described in the text
\label{fig:orb} }
\end{figure*}

\subsection{Rotational velocity}

The projected rotational velocity  ($v\sin{i}$) of this 
star has been previously estimated as $<12$ km s$^{-1}$ 
(Jeffries et al. 1995) and 18$\pm$2 km s$^{-1}$ by Cutisposto et al. (1999).


By using the program {\sc starmod} we have obtained the best fits 
for each observing run, 
with $v\sin{i}$ values around 15-18 km s$^{-1}$.
In order to determine a more accurate rotational velocity
of BK Psc we have made use of the cross-correlation technique 
in our high resolution echelle spectra by using the routine 
{\sc fxcor} in IRAF.
When a stellar spectrum with rotationally broadened lines is
cross-correlated against a narrow-lined spectrum, the width of the
cross-correlation function (CCF) is sensitive to the amount of 
rotational broadening of the first spectrum.
Thus, by measuring this width, one can obtain a measurement of 
the rotational velocity of the star.

The observed spectra of BK Psc were
cross-correlated against the spectrum of a template star
 (a slowly rotating star of similar spectral type) 
and the width (FWHM) of CCF determined. 
The calibration of this width
 to yield an estimation of $v\sin{i}$ is determined by
cross-correlating artificially broadened spectra of the template star
 with the original template star spectrum. 
The broadened spectra were created for $v\sin{i}$ spanning the
expected range of values by convolution with a theoretical rotational
profile (Gray 1992) using the program {\sc starmod}.
The resultant relationship between $v\sin{i}$ and FWHM of 
the CCF was fitted with a fourth-order polynomial.
 We have tested this method with stars of known rotational 
velocity, obtaining a good agreement. 
The uncertainties on the $v\sin{i}$ values obtained by this method
have been calculated using the parameter 
$R$ defined by Tonry \&  Davis (1979) 
as the ratio of the CCF height to the rms
antisymmetric component. This parameter is computed by the task {\sc fxcor}
and provides a measure of the signal to noise ratio of the CCF.
Tonry \&  Davis (1979) showed that errors in the FWHM 
of the CCF are proportional to $(1 + R)^{-1}$ and
Hartmann et al. (1986) and Rhode et al. (2001) found that the 
quantity $\pm v\sin{i}(1 + R)^{-1}$ provide a good estimate 
for the 90$\%$ confidence level of a $v\sin{i}$ measurement.
Thus, we have adopted   $\pm v\sin{i}(1 + R)^{-1}$ as a
reasonable estimate of the uncertainties on our $v\sin{i}$ measurements.

As template stars for BK Psc we have used the slowly rotating K5V stars 
 above mentioned in the spectral type classification.
We have determined $v\sin{i}$ by this method in all the spectra
of BK Psc available and the resulting weighted means in each 
observing run are 16.2$\pm$2.1, 16.4$\pm$0.7, 
19.5$\pm$0.8, and 15.0$\pm$1.0km s$^{-1}$ respectively.
The weighted mean for all the observing runs is 
17.1$\pm$0.5 km s$^{-1}$, which is the
value given in Table~\ref{tab:par}.

\begin{table}
\caption[]{Radial velocities of BK Psc
\label{tab:vr}}
\small
\begin{tabular}{llrrrrrrrrrrll}
\noalign{\smallskip}
\hline
\noalign{\smallskip}
 Obs. & HJD & \multicolumn{1}{c}{Primary} & & \multicolumn{1}{c}{Secondary} \\
\cline{3-3}\cline{5-5}
\noalign{\smallskip}
    & $2400000+$ & {\rm V$_{\rm hel}$} $\pm$ $\sigma_{\rm V}$ & &
                     {\rm V$_{\rm hel}$} $\pm$ $\sigma_{\rm V}$ \\
\noalign{\smallskip}
    &     & \small (km s$^{-1}$) & & \small (km s$^{-1}$) \\
\noalign{\smallskip}
\hline
\noalign{\smallskip}
 1992 (J95)$^{1}$ & 48845.581   &  2.9  $\pm$  2.0  & &  -        \\
 1992 (J95)$^{1}$ & 48851.540   & -56.9 $\pm$  2.0  & &  -        \\
\noalign{\smallskip}
 1995 (C99)$^{2}$ & 49956.718  & -26.3 $\pm$  4.0  & &   -               \\
 1995 (C99)$^{2}$ & 49958.774  & -40.6 $\pm$  4.0  & &   -               \\
\\ 
 2.2~m 1999 & 51385.615   & -29.80 $\pm$ 0.29 & &  31.92$\pm$4.3    \\
 2.2~m 1999 & 51387.650   &  -9.67 $\pm$ 0.46 & &  -9.67$\pm$4.5    \\
 2.2~m 1999 & 51389.633   &  16.51 $\pm$ 0.47 & & -57.65$\pm$4.5    \\
\noalign{\smallskip}
 INT 2000  & 51767.692   & -29.98 $\pm$ 0.13 & &  36.79$\pm$4.1    \\
 INT 2000  & 51768.689   &  22.71 $\pm$ 0.11 & & -70.23$\pm$4.1    \\
 INT 2000  & 51769.706   & -52.42 $\pm$ 0.13 & &  55.93$\pm$4.4    \\
 INT 2000  & 51770.703   &  37.91 $\pm$ 0.09 & & -93.27$\pm$4.1    \\
\noalign{\smallskip}
 NOT 2000  & 51854.551   &  -6.84 $\pm$ 0.21 & &  -6.84$\pm$4.2    \\
 NOT 2000  & 51855.470   &   3.06 $\pm$ 0.17 & & -46.55$\pm$4.2    \\
 NOT 2000  & 51856.527   & -32.54 $\pm$ 0.14 & &  31.00$\pm$4.1    \\
 NOT 2000  & 51857.486   &  27.93 $\pm$ 0.78 & & -75.80$\pm$4.8    \\
\noalign{\smallskip}
 2.2~m 2001 & 52176.554   & -56.65 $\pm$ 0.23 & &  71.82$\pm$4.2    \\
 2.2~m 2001 & 52177.448   &  16.25 $\pm$ 0.10 & & -57.80$\pm$4.1    \\
\noalign{\smallskip}
\hline
\noalign{\smallskip}
\end{tabular}

{
\small
$^{1}$ J95: Jeffries et al. (1995),
\\
$^{2}$ C99: Cutispoto et al. (1999).
}
\end{table}

\subsection{Radial velocities and orbital parameters}

Recently, Jeffries et al. (1995) and Cutispoto et al. (1999) 
have noted that BK Psc is a single-lined spectroscopic binary (SB1),
but they provided only four radial velocity measurements.
The detailed analysis of our spectra and our radial velocities measurements.
confirm the SB1 nature of this system.
In our spectra only the photospheric absorption lines coming from 
the primary component are observed throughout all the spectral range.
On the contrary, the chromospheric emission lines from both 
components are detected in our spectra (see Fig.~\ref{fig:ha}),
 and thereby it has been possible
to measure the radial velocity of the secondary and in this way 
to obtain the orbital solution of the system as in the case of a 
double-lined spectroscopic binary (SB2).

For the primary component of BK Psc the heliocentric radial velocities 
have been determined by using the cross-correlation technique.
The spectra of BK Psc were cross-correlated order by order, by
using the routine {\sc fxcor} in IRAF, against spectra of radial velocity
standards with similar spectral type taken from Beavers et al. (1979).
The radial velocity was derived for each order
from the position of the cross-correlation peak, 
and the uncertainties were calculated by {\sc fxcor} based on the
fitted peak height and the antisymmetric noise as described by
Tonry \&  Davis (1979).
In Table~\ref{tab:vr} we list, for each spectrum, the
heliocentric radial velocities ($V_{\rm hel}$)
and their associated errors ($\sigma_{V}$)
obtained as weighted means of the individual values deduced for each order.
Those orders that contain chromospheric features and prominent 
telluric lines have been excluded when determining the mean velocity.

For the secondary component, we have used the information provided by 
the chromospheric emissions that are detected for both components
in the H$\alpha$, Ca~{\sc ii} H \& K and other Balmer lines.
The contribution of each component to the observed profile 
has been deblended by mean of
a two Gaussian fit (see Fig.~\ref{fig:ha}) and the relative wavelength 
separation of the secondary component with respect to the primary 
has been used to determine
its heliocentric radial velocity (listed in Table~\ref{tab:vr}).

We have computed the orbital solution of BK Psc using our eleven 
data of radial velocities (for both component) 
and the four values given (only for the primary)
by Jeffries et al. (1995) and Cutispoto et al. (1999) 
(see Table~\ref{tab:vr}).
The radial velocity data (Table~\ref{tab:vr}) 
are plotted in Fig.~\ref{fig:orb}. 
Solid circles represent the primary and open circles represent the secondary.
The curves represent a minimum $\chi^{2}$ fit orbit solution.
The orbit fitting code uses the {\it Numerical Recipes} 
implementation of the Levenberg-Marquardt
method of fitting a non-linear function to the data and weights each datum
according to its associated uncertainty (Press et al. 1986). 
The program simultaneously solves for the 
orbital period, $P_{\rm orb}$, the epoch of periastron passage, $T_{0}$,
the longitude of periastron, $\omega$, the eccentricity, $e$, the primary 
star's radial velocity amplitude, $K_{1}$, 
the heliocentric center of mass velocity, $\gamma$, and the mass ratio, $q$. 
The secondary star's radial velocity amplitude, $K_{2}$, is $qK_{1}$.
The orbital solution and relevant derived quantities are given in
Table~\ref{tab:orb}.
In this Table we give $T_{\rm conj}$ as the heliocentric Julian date 
on conjunction
with the hotter star behind, in order to adopt the same criteria used
by Strassmeier et al. (1993) in their catalog of chromospherically
active binary stars.
We have used this criterion to calculate the orbital phases of all the 
observations reported in this paper.

\begin{table}
\caption[]{Orbital solution of BK Psc
\label{tab:orb}}
\small
\begin{tabular}{lccc}
\noalign{\smallskip}
\hline
\noalign{\smallskip}
Element & Value & Uncertainty & Units \\
\noalign{\smallskip}
\hline
\noalign{\smallskip}
 $P_{\rm orb}$      &  2.1663     & 0.0015  & \small days  \\
 $T_{\rm conj}$            &  2451383.32 & 0.20    & \small HJD  \\
 $\omega$           &  84.88      & 0.61    & \small degrees  \\
 $e$                &  0.0025     & 0.0074  &  \\
 $K_{1}$            &  52.70      & 0.73    & \small km~s$^{-1}$\\
 $K_{2}$            &  95.09      & 2.66    & \small km~s$^{-1}$  \\
 $\gamma$           & -10.95      & 0.32    & \small km~s$^{-1}$  \\
 $q=M_{1}/M_{2}$    &  1.80       & 0.04    &   \\
\\
 $a_{1}$~sin$i$     &  1.57       & 0.02    & \small 10$^{6}$~km \\
 $a_{2}~$sin$i$     &  2.83       & 0.08    & \small 10$^{6}$~km \\
 $a$~sin$i$         &  4.40       & 0.08    & \small 10$^{6}$~km \\
 "                  &  0.0294     &         & \small AU  \\
 "                  &  6.33       &         & \small R$_{\odot}$ \\
\\
 $M_{1}$~sin$^{3}i$ &  0.466      & 0.027   & \small $M_{\odot}$\\
 $M_{2}$~sin$^{3}i$ &  0.258      & 0.015   & \small $M_{\odot}$ \\
 $f(M)_{1}$         &  0.0328     & 0.0013  & \small $M_{\odot}$ \\
 $f(M)_{2}$         &  0.1917     & 0.0025  & \small $M_{\odot}$ \\
\noalign{\smallskip}
\hline
\end{tabular}
\end{table}

This binary system results in a 
circular orbit ($e$ = 0.0025) with an
 orbital period of 2.1663 days, which is
very similar to its rotational period derived from the photometry
($P_{\rm phot}$ = 2.24 days) indicating nearly synchronous rotation.


\subsection{Other derived quantities}


For the K5V spectral type of the observed primary component 
we can adopt from Landolt-B\"{o}rnstein tables (Schmidt-Kaler 1982) 
a primary mass $M_{1}$ = 0.67 $M_{\odot}$. 
According to the mass ratio ($q=M_{1}/M_{2}$ = 1.80) from the orbital
solution we estimate for the secondary a mass 
$M_{2}$ = 0.37 $M_{\odot}$ 
which corresponds (Schmidt-Kaler 1982) to a M3V star. 
With these spectral types 
the difference in bolometric magnitudes between
both components is 2.5 and
the difference in visual magnitudes is 
3.9 which is in agreement with an 
unseen secondary component and the spectral classification 
reported by Cutispoto et al. (1999). 


We have estimated the radius of the primary component by using the 
parallax (30.52 mas) given by Hipparcos (ESA 1997) and the unspotted 
$V$ magnitude, taken as the brightest magnitude (10.43) of the values 
given by Cutispoto et al. (1999). This $V$ magnitude is very close to 
the value given by  Hipparcos ($V_{T}=10.60$ that corresponds to $V=10.48$).
As the system is relatively close, to calculate the absolute magnitude 
$M_{V}$, no interstellar reddening was assumed.
The bolometric correction ($BC=-0.72$) 
corresponding to the K5V primary from (Schmidt-Kaler 1982) 
has been used to compute the bolometric magnitude, $M_{bol}$ 
and luminosity, $L/L_{\odot}$.
Assuming that the contribution of the secondary to this total 
luminosity is very small, we have used this $L/L_{\odot}$ and the
effective temperature ($T_{eff}=4350$ K) corresponding to a K5V 
to determine a radius (we called $R_{Hip}$) for the primary  
 $R_{Hip}=0.60\pm0.04 R_{\odot}$.
The errors in these derived quantities are dominated by the 
error in the parallax ($\pm1.79$ mas) given by Hipparcos and in 
the $T_{eff}$ ($\pm100$ K). 
This radius can be compared with an independent determination
of the minimum radius ($R\sin{i}$).
Taking as rotational period the photometric period (2.24 days) 
given by Cutispoto et al. (1999) and the rotational velocity 
 $v\sin{i}=17.1$ determined by us we found 
$R\sin{i}= 0.76\pm0.03 R_{\odot}$.
The error in this case is dominated by the uncertainty in $v\sin{i}$.
This value of $R\sin{i}$ should be smaller than $R_{Hip}$,
but we have found a value slightly larger.
Within the errors, however, the agreement between both radii is 
acceptable. In addition, the radius for a K5V in Schmidt-Kaler (1982),   
$R=0.72 R_{\odot}$, is halfway between $R_{Hip}$ and $R\sin{i}$.
This low value of $R_{Hip}$ also suggests that the $V$ magnitude 
we have used can be effected by cool spots 
on the stellar surface. 
Using the measured minimum radius $R\sin{i}$ and the 
effective temperature ($T_{eff}=4350$ K) corresponding to a K5V 
we obtain a low limit of the of the stellar luminosity ($L=0.167 L_{\odot}$) 
 and brightness ($V=9.88$).
Using this luminosity and the the mass-luminosity
relation for main sequence stars we obtain an estimate of the mass
of the primary (0.594 $M_{\odot}$) that is compatible with the mass of 
a K5V-K7V. With this mass and the  minimum mass
obtained from the orbital solution the mass for the secondary is 
$M_{2}$ = 0.33 $M_{\odot}$ which correspond to a M3V star,
similar to the result obtained with the first method described above.


We can summarize the adopted and derived quantities of the
primary and secondary components of BK Psc as follows:

\begin{center}
\[
Primary \ K5V = 
\left\{ 
\begin{array}{ll}
M_{1(K5V)}=0.67 , M_{1min}\approx0.59 \  M_{\odot} \\
R_{1(K5V)}=0.72 , R_{1}\sin{i}\approx0.76 \ R_{\odot} \\
T_{1(K5V)}=4350 \ K
\end{array} 
\right. \]
\end{center}

\begin{center}
\[
Secondary \  M3V = 
\left\{
\begin{array}{ll}
M_{2}\approx0.37 , M_{2min}\approx0.33 \ M_{\odot} \\
R_{2(M3V)}=0.45 \ R_{\odot} \\
T_{2(M3V)}=3470 \ K
\end{array} 
\right. \]
\end{center}

The inclination of the system is 
$i$ = 62.4$^{\small o}$ if we compare the minimum mass of the primary
($M_{1}$~sin$^{3}i$=0.466 $M_{\odot}$)
deduced from the orbit with the mass adopted for a K5V.
The minimum inclination angle for eclipses to occur ($i_{min}$) is 
given by $\cos{i_{min}}=(R_{1}+R_{2})/a$.
Using the radii adopted for the primary and secondary components and
the semi-mayor axis of the orbit ($a\sin{i}$) lead to 
$i_{min}=80.1^{\small o}$.
Since the photometric observations show no evidence 
of eclipses, the inclination of BK Psc must be lower than 
80.1$^{\small o}$, which is in agreement with the previous 
estimation of $i$.

\subsection{Kinematics and age}

BK Psc is a high proper-motion star included in the studies of
Stephenson (1986), Sandage \& Kowal (1986) and Weis (1991).
It is a relatively nearby star (d~=~32.8 pc) with astrometric 
data measured by Hipparcos (ESA 1997)
and Tycho-2 (H$\o$g et al. 2000) catalogues (see Table~\ref{tab:par}).

We have computed the galactic space-velocity components ($U$, $V$, $W$)
using as radial velocity the 
center of mass velocity ($\gamma$)
(for details see Montes et al. 2001b).
The resulting values and associated errors are given in Table~\ref{tab:uvw}.

\begin{table}[h]
\caption[h]{Galactic space-velocity components
\label{tab:uvw}}
\begin{flushleft}
\small
\begin{tabular}{l l l l l }
\noalign{\smallskip}
\hline
\noalign{\smallskip}
$U\pm \sigma_{U}$ & $V \pm \sigma_{V}$ & $W \pm \sigma_{W}$ & $V_{\rm Total}$ \\(km s$^{-1}$) & (km s$^{-1}$) & (km s$^{-1}$)  & (km s$^{-1}$) \\
\noalign{\smallskip}
\hline
\noalign{\smallskip}
 -55.23 $\pm$ 3.38  &  -63.27 $\pm$ 3.35  & -25.66 $\pm$ 2.04 & 87.82 \\
\noalign{\smallskip}
\hline
\noalign{\smallskip}
\end{tabular}
\end{flushleft}
\end{table}

The large total velocity $V_{\rm Total}$ and the $U$, $V$, $W$ 
velocity components that lie 
clearly outside the young disk population boundaries
in the  ($U$, $V$) and ($U$, $W$) diagrams 
(Eggen 1984, 1989; Montes et al. 2001b) indicate 
that BK Psc is an old disk star.


The spectral region of the resonance doublet 
of  Li~{\sc i} at $\lambda$6708 \AA\
is included in all our spectra of BK Psc.
The detailed analysis of the spectra indicates 
that this line is not detected in this star.
As it is well known, this spectroscopic feature 
is an important diagnostic of age in late-type stars. 
In addition, it is also known that a large number of
chromospherically active binaries
shows Li~{\sc i} abundances higher than
other stars of the same mass and evolutionary
stage (see Paper II, III and references therein). 
Therefore, non- detection of the Li~{\sc i} line in this active star 
indicates that it is an old star which is in agreement with its kinematics.

\begin{figure*}
{\psfig{figure=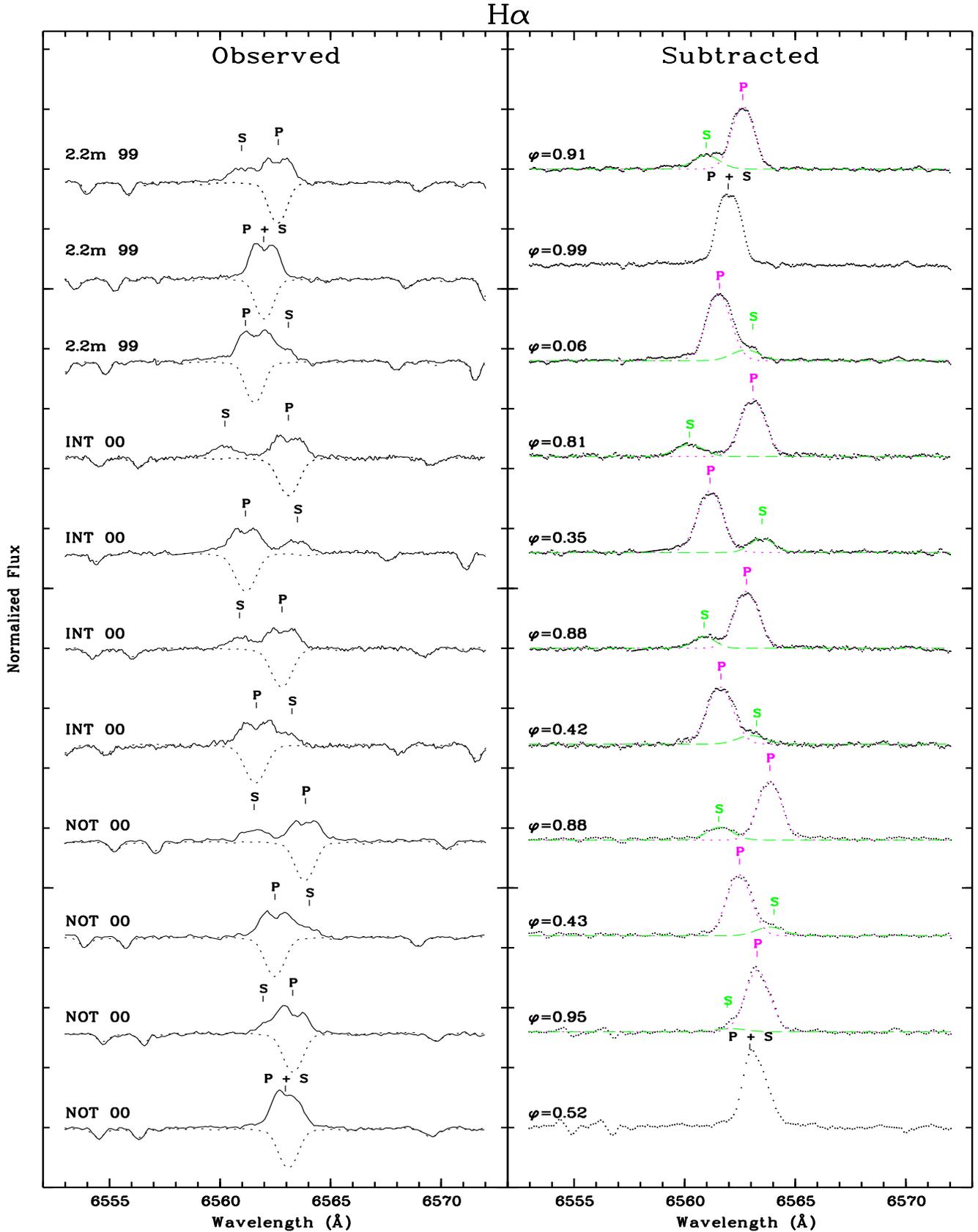,bbllx=32pt,bblly=32pt,bburx=547pt,bbury=777pt,height=22.5cm,width=17.8cm,clip=}}
\caption[ ]{Spectra of BK Psc in the H$\alpha$ line region. 
The observed spectrum (solid-line) and the
synthesized spectrum (dashed-line) are plotted in the left panel
and the subtracted spectrum (dotted line) in the right panel.
The position of the H$\alpha$ line for the primary (P) and secondary (S)
components are marked with short vertical lines.
We have superposed the two-Gaussian fit used to deblend, 
in the subtracted spectrum, the excess emission coming from both components.
\label{fig:ha} }
\end{figure*}

\section{Chromospheric activity indicators}

The echelle spectra analysed in this work allow us to
study the behaviour of the different indicators from
the Ca~{\sc ii} H \& K to the Ca~{\sc ii} IRT lines,
which are formed at different atmospheric heights.
The chromospheric contribution in these features has been 
determined by using the spectral subtraction technique described in
detail by Montes et al. (1995a) and Paper~I, II, and III.
The synthesized spectrum was constructed using the program {\sc starmod}.
Taken into account the stellar parameters derived in Sect.~3 we 
have used only a K5V primary component without any contribution from
a secondary component.
The inactive K5V stars used as reference stars are 
HD 154363 for the first observing run, 
and 61 Cyg A for the other three runs. 

In Table~\ref{tab:ew}  
we give the excess emission equivalent width ($EW$) (measured in the
subtracted spectra) for the Ca~{\sc ii} H \& K, H$\epsilon$,
H$\delta$, H$\gamma$, H$\beta$, H$\alpha$, and  Ca~{\sc ii} IRT
($\lambda$8498, $\lambda$8542, $\lambda$8662) lines.
 When the emission features from both components
can be deblended, we give the $EW$ for the hot and
cool (H/C) components. 
%
The uncertainties in the measured $EW$ were estimated taking into account:
a) the typical internal precisions of {\sc starmod}
(0.5 - 2 km s$^{-1}$ in velocity shifts, and
$\pm$5 km s$^{-1}$ in  $v\sin{i}$),
b) the rms obtained in the fit between observed and
synthesized spectra in the spectral regions outside the chromospheric features
(typically in the range 0.01-0.03)
 and 
c) the standard deviations resulting in the
$EW$ measurements. 
The final estimated errors are in the range 10-20\%.

The measured $EW$s given in Table~\ref{tab:ew} have been corrected for
the relative contribution of
each component to the total continuum
determined by means of the radii and
temperatures assumed in Sect.~3.
For instance, in the H$\alpha$ line region the relative contributions 
are $S_{\rm H}=0.94$ for the hot component and 
$S_{\rm C}=0.06$ for the cool component, and the corrected $EW$s
for the hot and cool components are obtained multiplying by a factor 
$1/S_{\rm H}$ and $1/S_{\rm C}$, respectively. 
Finally, these corrected $EW$s have been converted to
 absolute surface fluxes by using the
empirical stellar flux scales calibrated by Hall (1996)
as a function of the star colour index.
In our case, we have used the $B-V$ index and the corresponding coefficients 
for Ca~{\sc ii} H \& K, H$\alpha$ and Ca~{\sc ii} IRT, using the same as
Ca~{\sc ii} H \& K for H$\epsilon$, and derived the H$\delta$, H$\gamma$
and H$\beta$ fluxes by making an interpolation between the values of
Ca~{\sc ii} H \& K and H$\alpha$. 
The logarithm of the obtained absolute flux at the stellar surface
(logF$_{\rm S}$) for the different chromospheric activity indicators
is given in Table~\ref{tab:flux}.
  
\begin{figure}
{\psfig{figure=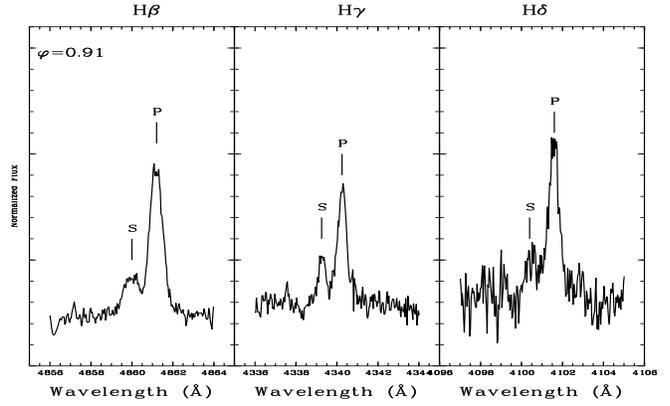,bbllx=27pt,bblly=37pt,bburx=550pt,bbury=527pt,height=5.3cm,width=8.6cm,clip=}}
\caption[ ]{Subtracted spectra in the region of the
H$\beta$, H$\gamma$, and H$\delta$ lines.
Clear excess emission from the primary (P) and secondary (S) components 
is detected
\label{fig:hbgd} }
\end{figure}

In Figs.~\ref{fig:ha} and \ref{fig:irt} we have plotted 
for each observation in the H$\alpha$ and 
Ca~{\sc ii} IRT $\lambda$$8498$, $\lambda$$8542$ line region
the observed spectrum (solid-line) 
and the synthesized spectrum (dashed-line) in the left panel,
and the subtracted spectrum (dotted line), in the right panel.
The observing run and the orbital phase ($\varphi$) of each spectrum
are also given in these figures.
The observed spectra in the Ca~{\sc ii} H \& K line region are 
plotted in Fig.\ref{fig:hyk}, and representative subtracted
spectra in the H$\beta$, H$\gamma$ and H$\delta$ line regions 
are plotted in Fig.~\ref{fig:hbgd}. 

\subsection{The H$\alpha$ line} 

The H$\alpha$ line region is included in our spectra in the four 
observing runs.
In all cases we have detected, in the observed spectra 
(see Fig.~\ref{fig:ha} left panel), strong H$\alpha$
emission above the continuum coming from the primary component
and a small H$\alpha$ emission coming from the secondary component.
In all the spectra, except two which are very close to conjunction, 
we were able to deblend the emission coming from both components 
by using a two-Gaussian fit to the subtracted spectra 
(see Fig.~\ref{fig:ha} right panel). 

The H$\alpha$ emission of the primary exhibits a central self-absorption
similar to that observed in many M active stars
(Stauffer \& Hartmann 1986) and some K active stars like 
the dK5e binary V833 Tau (Montes et. al. 1995b)
and the K4V single star V834 Tau (Montes et. al. 2001c).
This self-absorption feature is a consequence of the line formation process
in the chromosphere of very active stars (Houdebine \& Doyle 1994).

The H$\alpha$ emission of BK Psc is persistent during 
the period of time covered by our observations (from 1999 to 2001).
In addition, strong H$\alpha$ emission above the continuum from the 
primary component was also detected
in previous spectra of this system taken in 1992 with 
$EW$(H$\alpha$)~=~1.0 \AA\ (Jeffries et al. 1995) and 
$EW$(H$\alpha$)=1.1 \AA\ (Mason et al. 1995).
These $EW$s are lower than the $EW$s determined by us because these authors
determined the $EW$s in the observed spectra and our $EW$s have been measured
in the subtracted spectra, after eliminating the photospheric contribution. 
This persistent H$\alpha$ emission detected in BK Psc indicates that it is a 
very chromospherically active binary system 
similar to some RS CVn systems like V711 Tau, UX Ari, HU Vir, and DM UMa,
and some BY Dra systems like BY Dra itself, and  YY Gem, 
which always show H$\alpha$ emission above the continuum.

The detection of H$\alpha$ emission from the cool secondary 
component (M3V) of BK Psc indicates that this star has 
a very strong H$\alpha$ emission, 
since its photospheric contribution to the observed continuum
is practically negligible.
Strong H$\alpha$ emission is typical of the group of 
M type stars called dMe stars, some of which also show a
scaled-up version of solar flares and are known 
as flare stars of UV Cet type stars. These latter stars are
characterized by dramatic increases in the Hydrogen Balmer emission lines.
However, this seems not to be the case for the secondary component
of BK Psc since the H$\alpha$ emission we have detected has a similar
intensity in the four observing runs.

The H$\alpha$ emission of the primary component shows small variations
with the orbital phase, for instance, in the first run the $EW$ changes from 
1.7 to 1.1 \AA.
Seasonal variations are also detected, with larger $EW$(H$\alpha$) in 
1999 than in 2001.


\subsection{The H$\beta$, H$\gamma$ and H$\delta$ lines} 

The other three Balmer lines included in all our spectra
(H$\beta$, H$\gamma$ and H$\delta$) also show evidence of chromospheric
activity.
After applying the spectral subtraction, clear excess emission 
from both components is detected (see Fig.~\ref{fig:hbgd}).
When the S/N is high enough we have deblended the emission coming from 
both components by using a two-Gaussian fit to the subtracted spectra
(see Table~~\ref{tab:ew}).
The three lines show small seasonal and orbital phase variations
with the same trend that the H$\alpha$ line. 

We have also measured the ratio of
excess emission $EW$ in the H$\alpha$ and H$\beta$ lines,
$\frac{EW({\rm H\alpha})}{EW({\rm H\beta})}$, and the ratio
of excess emission $\frac{E_{\rm H\alpha}}{E_{\rm H\beta}}$
with the correction:

\[ \frac{E_{\rm H\alpha}}{E_{\rm H\beta}} =
\frac{EW({\rm H\alpha})}{EW({\rm H\beta})}*0.2444*2.512^{(B-R)}\]
given by Hall \& Ramsey (1992) that takes into account 
the absolute flux density in these lines and the color difference 
in the components.
We have obtained for the primary component 
$\frac{E_{\rm H\alpha}}{E_{\rm H\beta}}$ in the range 
of 3 to 4 in all our spectra. These values indicate,
according to Buzasi (1989) and Hall \& Ramsey (1992),
the presence of prominence-like material at the stellar
surface.

\begin{figure}
{\psfig{figure=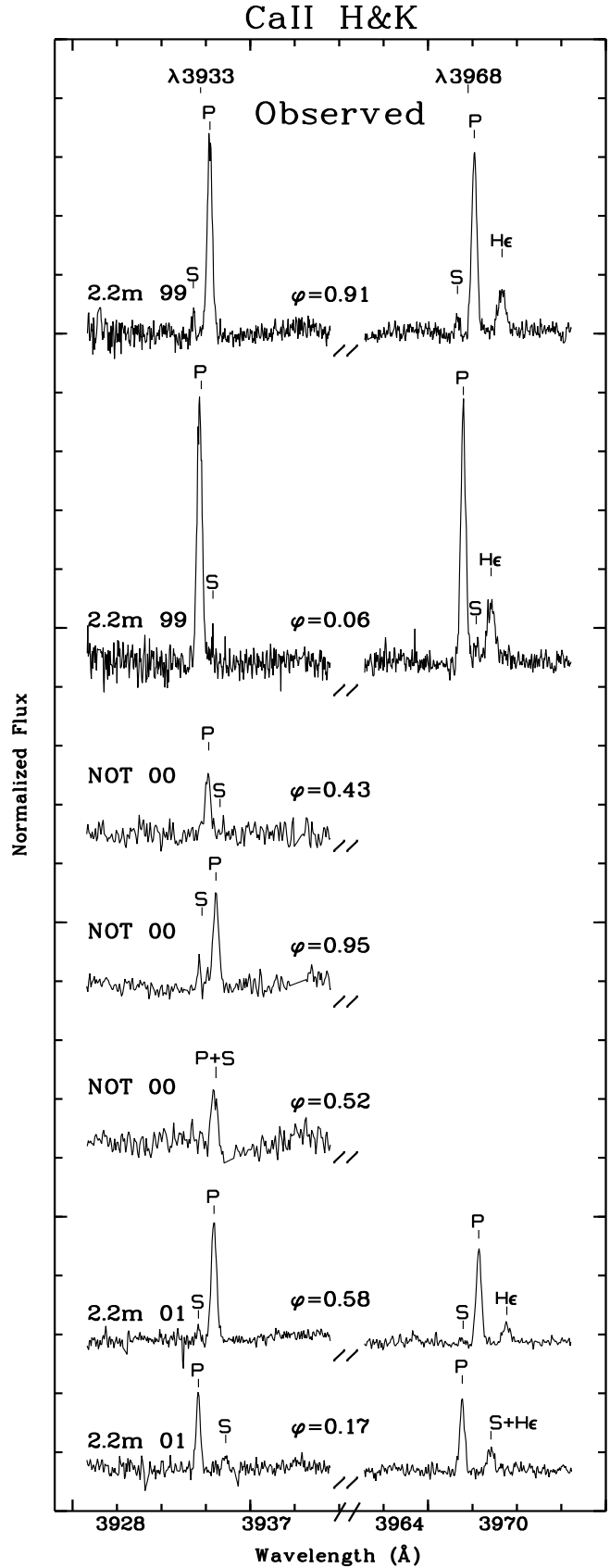,bbllx=36pt,bblly=35pt,bburx=301pt,bbury=775pt,height=22.5cm,width=8.6cm,clip=}}
\caption[ ]{Observed spectra in the region of the
Ca~{\sc ii} H \& K and H$\epsilon$ lines.
The position of these lines for the primary (P) and secondary (S)
components are marked with short vertical lines
\label{fig:hyk} }
\end{figure}

\begin{figure*}
{\psfig{figure=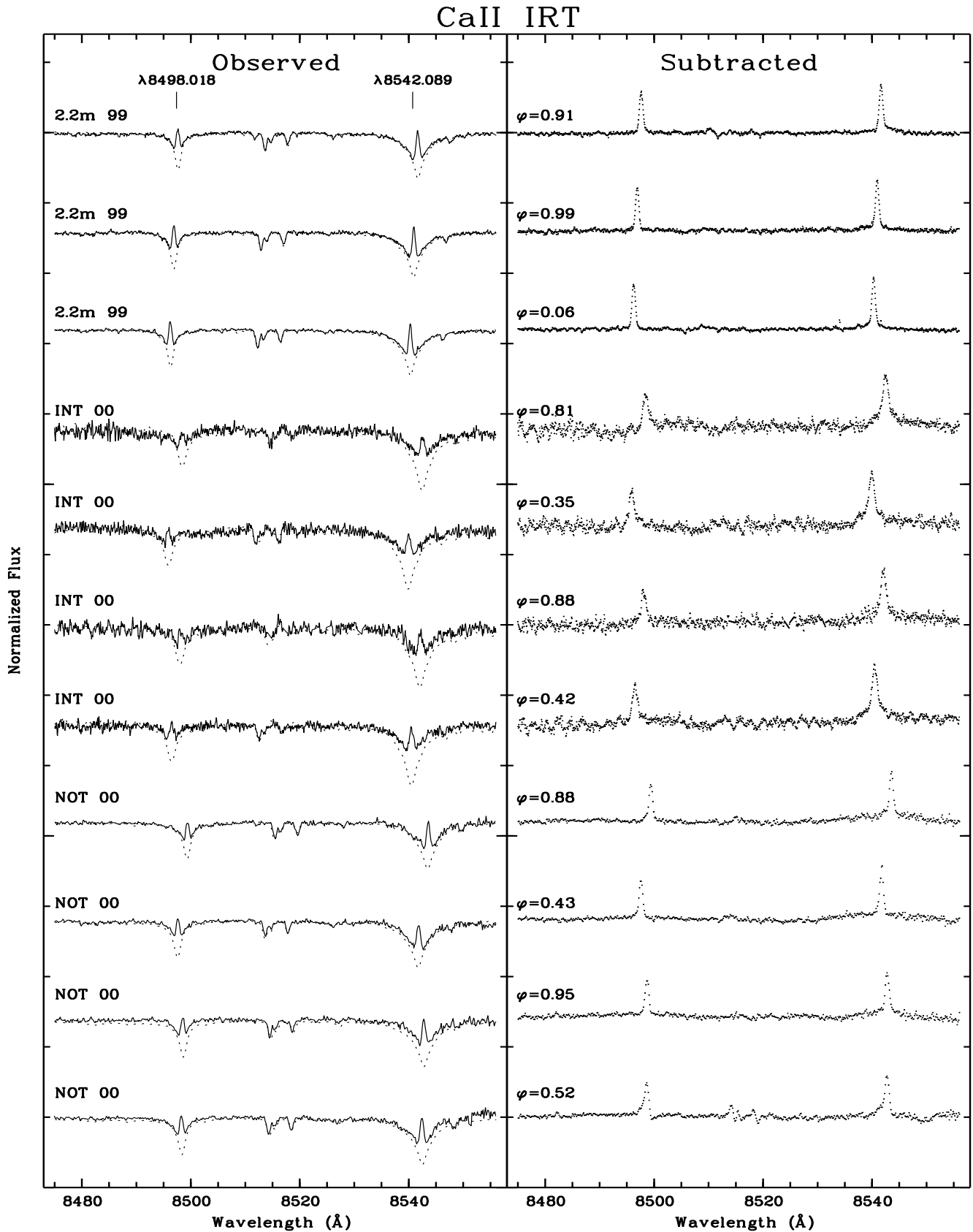,bbllx=32pt,bblly=32pt,bburx=547pt,bbury=777pt,height=22.5cm,width=17.8cm,clip=}}
\caption[ ]{Observed and subtracted spectra,
as in Fig.~\ref{fig:ha}, in the region of the
Ca~{\sc ii} IRT (8498, 8542~\AA) lines
\label{fig:irt} }
\end{figure*}

\subsection{The Ca~{\sc ii} H \& K and H$\epsilon$ lines} 

The Ca~{\sc ii} H \& K line region is included in the spectra
of the FOCES 1999 and 2001 observing runs. 
Only the Ca~{\sc ii} K line is included in the NOT 2000 run, 
and this spectral region is not covered in the MUSICOS 2000 run.
In all the spectra strong emission in the Ca~{\sc ii} H \& K 
lines and a clear emission in the H$\epsilon$ line
coming from the primary component is observed (see Fig.~\ref{fig:hyk}).
The Ca~{\sc ii} H \& K emission lines from the secondary 
component are also detected with relative wavelength shifts with
respect to the primary in agreement with the 
wavelength shifts calculated in the H$\alpha$ 
and the other Balmer lines.

In our spectra the Ca~{\sc ii} H \& K line spectral region 
is located at the end of the echellogram, where the efficiency of
the spectrograph and the CCD decrease very rapidly, and therefore the 
 S/N ratio obtained is very low, and the normalization of the spectra 
is very difficult. 
For these reasons we have not applied the spectral subtraction in 
this spectral region, and we have plotted in Fig.~\ref{fig:hyk} 
only the observed spectra.
The $EW$s measured in these spectra are in agreement with the 
strong Ca~{\sc ii} K emission (EW=~2.7 \AA) 
reported by Mason et al. (1995).
 
Variations of the Ca~{\sc ii} H \& K emissions with the 
orbital phase and from one  epoch to another are observed
for the primary component, 
and as in the case of the H$\alpha$ line the level 
of chromospheric activity is higher in the first observing run
than in the last one.



\subsection{The Ca~{\sc ii} IRT lines 
($\lambda$8498, $\lambda$8542, $\lambda$8662)} 

The three lines of the Ca~{\sc ii} infrared triplet (IRT) are
included in our echelle spectra, except $\lambda$8662 which is 
not included in the NOT 2000 observing run.
In all the observed spectra of BK Psc a clear emission reversal 
is observed in the core of the Ca {\sc ii} IRT absorption lines
(see Fig.~\ref{fig:irt} left panel).
After applying the spectral subtraction technique, only the 
excess emission arising from the primary component is detected
(see Fig.~\ref{fig:irt} right panel).
Unlike the other chromospheric activity indicators, no evidence 
of excess emission is detected from the cooler secondary component,
even when in this red spectral region its relative contribution 
is slightly larger than in the blue one.
In addition, the orbital phase and seasonal variations detected in the
Ca {\sc ii} IRT excess emission are more smooth than in the other
activity indicators. 


We have calculated the ratio of excess emission $EW$,
$\frac{E_{8542}}{E_{8498}}$, which is also an indicator of the
type of chromospheric structure that produces the observed emission.
In solar plages, values of
$\frac{E_{8542}}{E_{8498}}$ $\approx$~1.5-3 are measured,
while in solar prominences the values are $\approx$~9, 
the limit of an optically thin emitting plasma (Chester 1991).
The small values of the $\frac{E_{8542}}{E_{8498}}$ ratio we have found
for the primary component in all our spectra (ranging from 1.3 to 2.4)
indicate that the Ca~{\sc ii} IRT emission of this star 
arises from plage-like regions, in contrast to the Balmer lines
that seem to come from prominences.
This markedly different behaviour of the Ca~{\sc ii} IRT emission 
has also been found in other chromospherically
active binaries (see Paper~III and references therein).


\section{Conclusions}

In this paper a detailed spectroscopic analysis of the recently,
X-ray/EUV selected chromospherically active binary BK Psc is presented.
We have analysed high resolution echelle spectra taken during 
four observing runs from 1999 to 2001.
These spectra include all the optical chromospheric activity indicators
from the Ca~{\sc ii} H \& K to Ca~{\sc ii} IRT lines, as well as the 
 Li~{\sc i} $\lambda$6707.8 line and other photospheric lines of interest.

The precise radial velocities of the primary component 
of BK Psc that we have determined, by using the cross-correlation technique,
allow us to confirm that it is a single-lined spectroscopic binary (SB1).
In addition, as the chromospheric emission lines from the
 secondary component are also detected in our spectra it has been possible
to measure the radial velocities of the secondary.
We have used all these radial velocities to obtain, for the first time, 
the orbital solution of the system, as in the case of a SB2 system.
We have obtained a circular orbit with an orbital period 
of 2.1663 days, very close to its photometric period of 2.24 days 
(indicating synchronous rotation). 
By comparison with the spectra of reference stars of different spectral types
we have found that the spectra of BK Psc is well fitted with a K5V primary 
component without any contribution of a secondary component.
The minimum masses ($Msin^{3}i$) resulting from the orbital 
solution are compatible with the observed K5V primary 
and an unseen M3V secondary.
As both components are main sequence stars, we can classify this
chromospherically active binary as a BY$~$Dra system 
(Fekel et al. 1986). 

By using the information provided 
by the width of the cross-correlation function 
we have determined a projected rotational
velocity, $v\sin{i}$, of 17.1 km s$^{-1}$ for the primary component.
Within the errors, the radius for a K5V ($R=0.72 R_{\odot}$) is 
compatible with the radius obtained from the  {\sc Hipparcos} data
($R_{Hip}=0.60\pm0.04 R_{\odot}$) and the minimum radius 
($R\sin{i}=0.76\pm0.03 R_{\odot}$) obtained from $v\sin{i}$ and
the photometric period.
 
The kinematics ($U$, $V$, $W$ space-velocity components)
and the non- detection of the Li~{\sc i} line in this active star indicate
that it is an old star.

Finally, we have analysed, using the spectral subtraction technique,
all the optical chromospheric activity indicators
from the Ca~{\sc ii} H \& K to Ca~{\sc ii} IRT lines.
Both components of the binary system show 
high levels of chromospheric activity.
In the observed spectra we have detected strong emission in the H$\alpha$, 
Ca~{\sc ii} H \& K, H$\epsilon$, and  Ca~{\sc ii} IRT lines   
coming from the primary component, 
and low emission in the H$\alpha$ and Ca~{\sc ii} H \& K lines from the 
secondary.
After applying the spectral subtraction the chromospheric excess emission
from both components are clearly detected in all the activity indicators
except the Ca~{\sc ii} IRT lines.
We have used a two-Gaussian fit to deblend the emission coming from 
both components.
The H$\alpha$ emission above the continuum from both components 
seems to be a persistent feature of this system during
the period 1999 to 2001 of our observations as well as in previous
observations reported by other authors.    
In addition, the H$\alpha$ emission of the primary exhibits 
a central self-absorption.
The ratio $\frac{E_{\rm H\alpha}}{E_{\rm H\beta}}$
that we have found indicates that the emission of these lines
could arise from prominence-like material,
whereas the ratio $\frac{E_{8542}}{E_{8498}}$ 
indicates that the Ca~{\sc ii}~IRT emission 
arises from plage-like regions.
The excess emission $EW$ of the Balmer lines and the Ca~{\sc ii} H \& K lines
correlate well with each other, but the Ca~{\sc ii} IRT lines 
show a different behaviour.




\begin{table*}
\caption[]{$EW$ of the different chromospheric activity indicators of BK Psc
\label{tab:ew}}
\begin{flushleft}
\scriptsize
\begin{tabular}{cccccccccccc}
\noalign{\smallskip}
\hline
\noalign{\smallskip}
        &     & \multicolumn{10}{c}{$EW$(\AA) in the subtracted spectrum} \\
\cline{3-12}
\noalign{\smallskip}
 Obs. & $\varphi$$^{\star}$ & \multicolumn{2}{c}{CaII} & & & & & &
\multicolumn{3}{c}{CaII IRT} \\
\cline{3-4}\cline{10-12}
\noalign{\smallskip}
     &     &
 K$^{3}$   & H$^{3}$  & H$\epsilon$$^{3}$ & H$\delta$ & H$\gamma$ & H$\beta$ & H$\alpha$ &
\scriptsize $\lambda$8498 & \scriptsize $\lambda$8542 & \scriptsize $\lambda$8662
\small
\\
\noalign{\smallskip}
\hline
\noalign{\smallskip}

 2.2~m 99 & 0.06 &
6.14/0.43 & 7.03/0.68 & 1.73 & 0.52 & 0.58 & 0.73$^{1}$ & 1.70/0.14 & 0.55 & 0.70 & 0.63 \\
 2.2~m 99 & 0.99 &
  *  &   *  &   *  & 0.29 & 0.49 & 0.66$^{1}$ & 1.66$^{1}$ & 0.49 & 0.65 & 0.61 \\
 2.2~m 99 & 0.91 &
4.54/0.38 & 3.78/0.26 & 1.31 & 0.52/0.17 & 0.31/0.10 & 0.64/0.17 & 1.32/0.35 & 0.49 & 0.69 & 0.83 \\
\noalign{\smallskip}
 INT 00  & 0.42 &
  -  &   -  &  -   &  *   & 0.17+& 0.58$^{1}$ & 1.51/0.12 & 0.67 & 1.62 & 1.30 \\
 INT 00  & 0.88 &
  -  &   -  &  -   &  *   & 0.27 & 0.50/0.05  & 1.32/0.24 & 0.59 & 1.36 & 1.09 \\
 INT 00  & 0.35 &
  -  &   -  &  -   &  *   & 0.28 & 0.51/0.01  & 1.48/0.31 & 0.69 & 1.24 & 1.35 \\
 INT 00  & 0.81 &
  -  &   -  &  -   &  *   &  *   & 0.54/0.08 & 1.37/0.32 & 0.64 & 1.36 & 1.42 \\
\noalign{\smallskip}
 NOT 00  & 0.52 &
2.04 &   -  &  -   &  *   &0.098+& 0.73$^{1}$ & 1.79$^{1}$ & 0.55 & 0.76 &  -   \\
 NOT 00  & 0.95 &
3.56/0.38 &   -  &  -   &  *   &  *   & 0.60$^{2}$ & 1.46/0.05 & 0.59 & 0.90 &  -   \\
 NOT 00  & 0.43 &
2.15 &   -  &  -   &  *   &  *   & 0.57$^{2}$ & 1.49/0.16 & 0.59 & 0.75 &  -   \\
 NOT 00  & 0.88 &
 *   &   -  &  -   &  *   &0.246+& 0.55/1.13 & 1.32/0.31 & 0.53 & 0.89 &  -   \\
\noalign{\smallskip}
 2.2~m 01  & 0.17 &
1.88/0.37 & 2.31/$^{4}$ & 1.01$^{4}$ & 0.50/0.35 & 0.50/0.13 & 0.45/0.04 & 1.18/0.32 & 0.51 & 0.65 & 0.48  \\
 2.2~m 01  & 0.58 &
3.57/0.31+& 3.17/0.14+& 0.91 & 0.27/0.03 & 0.36/0.12 & 0.52/0.11 & 1.09/0.37 & 0.46 & 0.64 & 0.51  \\
\noalign{\smallskip}
\hline
\end{tabular}
\end{flushleft}
{
\small
$^{\star}$ Orbital phase calculated with the orbital period,
P$_{\rm orb}$ and date of conjunction, T$_{\rm conj}$ determined
in this paper.
\\
$*$ Data not measured due the very low S/N.
\\
+ Data measured with low S/N.
\\
$^{1}$ Data for the primary and secondary components not deblended due
to the orbital phase of the observation. \\
$^{2}$ For these lines we can observe the two components but they could not be
deblended.\\
$^{3}$ These values have been measured at the observed spectra.\\ 
$^{4}$ The H$\epsilon$ line of the primary component is blended with
the Ca~{\sc ii} H line of the secondary. \\
}
\end{table*}

\begin{table*}
\caption[]{Absolute surface flux
of the different chromospheric activity indicators of BK Psc
\label{tab:flux}}
\begin{flushleft}
\scriptsize
\begin{tabular}{cccccccccccc}
\noalign{\smallskip}
\hline
\noalign{\smallskip}
      &     & \multicolumn{10}{c}{logF$_{\rm S}$ in the subtracted spectrum} \\
\cline{3-12}
\noalign{\smallskip}
 Obs. & $\varphi$$^{\star}$ & \multicolumn{2}{c}{CaII} & & & & & &
\multicolumn{3}{c}{CaII IRT} \\
\cline{3-4}\cline{10-12}
\noalign{\smallskip}
     &     &
 K$^{3}$   & H$^{3}$  & H$\epsilon$$^{3}$ & H$\delta$ & H$\gamma$ & H$\beta$ & H$\alpha$ &
\scriptsize $\lambda$8498 & \scriptsize $\lambda$8542 & \scriptsize $\lambda$8662
\small
\\
\noalign{\smallskip}
\hline
\noalign{\smallskip}
2.2~m 99 & 0.06 &
6.75/5.48 & 6.81/5.69 & 6.20 & 5.70 & 5.68 & 5.70$^{1}$ & 6.55/6.14 & 6.16 & 6.26 & 6.22 \\
2.2~m 99 & 0.99 &
  *  &   *  &   *  & 5.44 & 5.71 & 5.93$^{1}$ & 6.54$^{1}$ & 6.11 & 6.23 & 6.20 \\
2.2~m 99 & 0.91 &
6.60/5.42 & 6.52/5.26 & 6.06 & 5.70/5.09 & 5.51/4,91 & 5.91/5.24 & 6.44/6.53 & 6.11 & 6.26 & 6.39 \\
\noalign{\smallskip}
INT 00  & 0.42 &
  -  &   -  &  -   &  *   & 5.25+& 5.87$^{1}$ & 6.50/6.07 & 6.24 & 6.63 & 6.53 \\
INT 00  & 0.88 &
  -  &   -  &  -   &  *   & 5.45 & 5.80/4.69 & 6.44/6.37 & 6.19 & 6.55 & 6.45 \\
INT 00  & 0.35 &
  -  &   -  &  -   &  *   & 5.47 & 5.81/3.86 & 6.50/6.48 & 6.26 & 6.51 & 6.55 \\
INT 00  & 0.81 &
  -  &   -  &  -   &  *   &  *   & 5.84/4.91 & 6.46/6.50 & 6.22 & 6.55 & 6.57 \\
\noalign{\smallskip}
NOT 00  & 0.52 &
6.26 &   -  &  -   &  *   & 5.01+& 5.97$^{1}$ & 6.55$^{1}$ & 6.16 & 6.30 &  -   \\
NOT 00  & 0.95 &
6.50/5.42 &   -  &  -   &  *   &  *   & 5.88$^{2}$ & 6.48/5.71 & 6.19 & 6.37 &  -   \\
NOT 00  & 0.43 &
6.28 &   -  &  -   &  *   &  *   & 5.86$^{2}$ & 6.49/6.20 & 6.19 & 6.29 &  -   \\
NOT 00  & 0.88 &
 *   &   -  &  -   &  *   & 5.41+& 5.85/6.07 & 6.44/6.48 & 6.14 & 6.37 &  -   \\
\noalign{\smallskip}
2.2~m 01 & 0.17 &
6.22/5.41 & 6.31/$^{4}$ & 5.95$^{4}$ & 5.68/5.41& 5.72/5.02 & 5.76/4.62 & 6.39/6.50 & 6.13 & 6.23 & 6.10 \\
2.2~m 01 & 0.58 &
6.50/5.34+& 6.44/4.99+& 5.91 & 5.41/4.34 & 5.58/4.98 & 5.82/5.05 & 6.39/6.56 & 6.08 & 6.22 & 6.13 \\
\noalign{\smallskip}
\hline
\end{tabular}
\end{flushleft}
{
\small
Notes as in Table~\ref{tab:ew}.
}
\end{table*}



\begin{acknowledgements}

We would like to thank Dr. B.H. Foing for allow us to use the
ESA-MUSICOS spectrograph at Isaac Newton Telescope.
We would also like to thank the referee S.V. Berdyugina
for suggesting several improvements and clarifications.
This work has been supported by the Universidad Complutense de Madrid
and the Spanish  MCYT, Ministerio de Ciencia y Tecnolog\'{\i}a, 
Programa Nacional de Astronom\'{\i}a y Astrof\'{\i}sica,
under grant AYA2001-1448.

\end{acknowledgements}



\end{document}